\documentclass[12pt]{article}
\usepackage[english]{babel}      
\usepackage[T1]{fontenc}       
\begin{document}
\pagenumbering{arabic}
\begin{titlepage}
\title{\normalsize{Gravitational Pressure and the Accelerating Universe}}
\author{{\footnotesize K. H. C. Castello-Branco $\, ^{1\ast}$ and 
J. F. da Rocha-Neto $\, ^{2 \dagger}$}}
\date{}
\maketitle

\begin{center}
{\footnotesize 1  Universidade Federal do Oeste do Par\'a,
Av. Marechal Rondon, 68040-070. Santar\'em, PA, Brazil.\\
2 Instituto de F\'isica, Universidade de Bras\'ilia, 
70910-900, Bras\'ilia, DF, Brazil.}
\end{center}

\begin{abstract}

\footnotesize{In the context of the Teleparallel Equivalent of General Relativity (TEGR) 
one can obtain an alternative insight into General Relativity, as has been shown 
in addressing properties as energy, momentum and angular momentum of the gravitational field. 
In this paper, we apply the definition, that arises from the field equation of the the TEGR, 
for the stress-energy-momen-\linebreak tum tensor of the gravitational field, whose spatial components naturally lead to the 
definition of \textit{gravitational pressure}, to compute the total space-time pressure, 
due to the gravitational and matter fields, over a spherical, space-like two-surface of a Friedman-Robertson-Walker (FRW) universe, 
for any curvature index. In particular, for a spatially flat FRW universe in the actual era (\textit{i.e.}, for 
``cold matter''), it resulted that the pressure (now due only to the gravitational field) is \textit{outwardly} directed over any spherical, 
spatial two-surface. This surface can be, in particular, the apparent horizon of a spatially flat FRW universe 
(in this case, the apparent horizon coincides with the Hubble horizon). Assuming the validity of the first law of thermodynamics 
for matter and gravity, and taking into account the contribution of the gravitational field to both the energy and the pressure 
terms in the first law of the (gravitational) thermodynamics, as well as considering the thermal character of the apparent horizon 
of the spatially flat FRW universe, we have thus obtained a value of the gravitational pressure that is very close to the observed 
value. We interpret this result as a possibility that the accelerated expansion of the actual universe might be due to the effect 
of the pressure of the very gravitational field, instead of an totally unkown (dark) energy.} 

\end{abstract}
 
{\footnotesize Keywords: {\it stress-energy-momentum of the gravitational and matter fields, gravitational pressure}

PACS numbers: 04.20.-q, 04.20.Cv, 04.70.Dy}

{\footnotesize
\noindent $\ast$ khccb@yahoo.com.br
\noindent $\dagger$ rocha@fis.unb.br}

\end{titlepage}

\section{Introduction}

\noindent

Direct evidence from precise measurements of the cosmological distance-redshift relation of type-Ia supernovae 
has indicated with high confidence that the expansion of the universe is accelerating \cite{Ref01,Perlmutter,Riess-Schmidt}. 
The data has also been interpreted as an evidence for a universe endowed with a positive ``vacuum energy'' density, which has usually 
been identified with the cosmological constant (widely interpreted as ``vacuum energy''), whose origin is totally unknown, and which 
dominates, as compared to matter, the present expansion of the universe. That sort of energy has been called \textit{``dark energy''} 
(see, \textit{e.g.}, \cite{DE}). Evidence that dark energy would contribute with a large fraction 
to the energy content of the universe has also been given by measurements of the cosmic microwave background (see, \textit{e.g.}, \cite{Dodelson}). 
While dark energy has been invoked to account for observational data, 
its nature and origin remain totally unexplained. 
Besides the cosmological constant, the option for a varying dark energy (the so-called quintessence), 
as described by a dynamical scalar field, has been widely studied (for a list of references on this as well as other proposals 
for describing dark energy see, for instance, \cite{Linder-AJP}) as an the entity responsible for cosmic speed up. 
Nevertheless, as the origin of such a scalar field is not known 
and in general fine-tunnings must be invoked to justify why such field has not yet been detected \cite{Carroll}, 
this is a largely \textit{ad hoc} description of dark energy. There have also been  proposals that the cosmic acceleration has 
its origin in modifications of General Relativity (see, for instance, \cite{Carroll-Trodden-Turner}). In spite of that, the theoretical 
explanation of the observed cosmic speeding up remains a mystery and it constitutes a challenge in physics since its discovery.

\bigskip

In this work, we consider the role that the \textit{pressure of the gravitational field} itself might have in addressing the issue of the accelerated expansion of the present universe. In order to address it we apply, to a Friedman-Robertson-Walker (FRW) universe endowed with a cosmological constant, a definition for the \textit{gravitational stress-energy-momentum} which naturally arises in the context of the \textit{Teleparallel Equivalent 
of General Relativity} (TEGR) \cite{Hehl1,Nester1989,Maluf1994,Hehl2,Maluf4,Obukov-Pereira,Obukov-Rub,Nester,
Maluf-Ulhoa2008,CastelloBr-Rocha2012,Aldrovandi-JG}. The TEGR is an alternative 
geometrical description to General Relativity, based on torsion and tetrad fields, rather than on curvature and metric. However, as indicated 
by its name, this theory is \textit{equivalent} to General Relativity, since the field equation of the TEGR reduces to the 
Einstein field equation, given in terms of tetrad fields. 
For a review on the TEGR, see,\textit{ e.g.}, Ref. \cite{ReviewTEGR-Maluf}. 

\bigskip

In the framework of the TEGR, the concept of gravitational stress-energy-momentum 
has been shown to be well-defined \cite{Maluf4, CastelloBr-Rocha2012, Maluf-Faria-CB, Ref16, Ref18, Ref19, Ref20} 
and from it the notion of \textit{gravitational pressure} has also been  
shown to naturally arise \cite{Ref18, Ref19}. Therefore, as it happens 
with the electromagnetic field, the gravitational field is also endowed with \textit{pressure}. Such definition of gravitational pressure has been 
applied in the context of gravitational waves \cite{Maluf-Ulhoa2008}, of a Schwarzschild black hole \cite{Ref18}, and, more recently, 
for Kerr \cite{Ref19} and Reissner-Nordstr\"om \cite{karl-jf} black holes. In all those applications, the definition of gravitational 
pressure has led to physically consistent results.

%
%
\bigskip

In the present work, we have computed the total pressure, due to gravitational and matter fields, over a spacelike spherical surface in a 
general (\textit{i.e.}, for any curvature index) Friedman-Robertson-Walker universe, endowed with a cosmological constant, in the framework 
of the TEGR. We have then applied it to a spatially flat universe, in order to address how the gravitational pressure could affect the expansion of the present universe. 


\bigskip

We set our notation as the following: space-time indices $\mu, \nu, ...$ and $SO(3,1)$ indices $a, b, ...$ rum from 0 to 3. 
Time and space indices are indicated according to $\mu =0, i$ and $a = (0),(i)$. Tetrad fields are denoted by $e^{a}\,_{\mu}\,$, 
and the torsion tensor reads $T_{a\mu\nu} = \partial_{\mu}e_{a\nu} - \partial_{\nu}e_{a\mu}$. The flat, Minkowski 
space-time metric tensor raises and lowers tetrad indices and is fixed by $\eta_{ab} = 
(-1, +1, +1, +1)$. 

\bigskip

\section{The Gravitational Energy-Momentum and \\ 
Pressure in the TEGR}

\noindent

In this section we present a summary of the Teleparallel Equivalent of General Relativity (TEGR) \cite{Hehl1}-\cite{Aldrovandi-JG} and the definition of the gravitational  stress-energy-momentum, as well as the total energy-momentum of gravitational and matter fields, 
which naturally arise from the field equation of the TEGR, when the latter is 
rewritten as a continuity equation \cite{ReviewTEGR-Maluf,Ref18}. 

\bigskip

In the TEGR the basic variables for the description of  the gravitational field are \textit{tetrad fields}, $e^{a}\,_{\mu}$, 
and the Lagrangian density is written in terms
of a quadratic combination of the torsion tensor, $T_{a\mu\nu}$, which is related
to the antisymmetric part of the Weitzenb\"ock connection, $\Gamma^{\lambda}\,_{\mu\nu}=
e^{a\lambda}\partial_{\mu}e_{a\nu}$. The Riemann curvature tensor corresponding to this connection identically vanishes and, thus, 
one has a space-time manifold with distant (or absolute) parallelism, \textit{i.e.}, with \textit{teleparallelism}. 
In the TEGR the Lagrangian density for the gravitational and matter fields is defined as (see Ref. \cite{ReviewTEGR-Maluf} and references therein)

\begin{eqnarray}
L&=& -\kappa e\left(\frac{1}{4}T^{abc}T_{abc}+\frac{1}{2}T^{abc}T_{bac}-
T^aT_a\right) - \frac{1}{c}L_m\nonumber \\
&\equiv& -\kappa e\Sigma^{abc}T_{abc}-\frac{1}{c}L_{m}\,, 
\label{2.2}
\end{eqnarray}
in which $\kappa\equiv c^{3}/16\pi G\,$, $e\equiv det(e^{a}\,_{\mu})$, $T_{abc}=e_b\,^\mu e_c\,^\nu T_{a\mu\nu}$, $T_a=T^b\,_{ba}\,$,
and $\Sigma^{abc}$ is defined by 

\begin{equation}
\Sigma^{abc}= \frac{1}{4}\left(T^{abc}+T^{bac}-T^{cab}\right)
+\frac{1}{2}\left(\eta^{ac}T^b-\eta^{ab}T^c \right)\,,
\label{2.3}
\end{equation}
and $L_m$  is the Lagrangian density for matter fields. 
The variation of $L$ with respect to the tetrad field $e^{a\mu}$ gives the field equation of the TEGR \cite{Maluf1994} (see also 
the review \cite{ReviewTEGR-Maluf})
\begin{equation}
e_{a\lambda}e_{b\mu}\partial_\nu (e\Sigma^{b\lambda \nu} )-
e \left(\Sigma^{b\nu}\,_aT_{b\nu\mu}-
\frac{1}{4}e_{a\mu}T_{bcd}\Sigma^{bcd}\right)=\frac{1}{4\kappa c}eT_{a\mu}\,.
\label{2.4}
\end{equation}
We note that the stress-energy-momentum of matter fields is given by $T^{\lambda\mu}=e_a\,^{\lambda}T^{a\mu}\,$, which is defined through $eT_{a\mu}=\delta L_m / \delta e^{a\mu}$. 
The left side of the above equation may be identically rewritten in terms of the left side of Einstein equation, as 
$\frac{1}{2}e\left[ R_{a\mu}(e)-\frac{1}{2}e_{a\mu}R(e)\right]$ (for details, see Ref. \cite{ReviewTEGR-Maluf}). We remark 
that as the Lorentz indices in Eq. (\ref{2.4}) may be converted into space-time indices, then the left side of the latter 
becomes proportional to $(R_{\mu\nu} - \frac{1}{2}g_{\mu\nu}R)$. This implies that a metric tensor, $g_{\mu\nu}$, that is a solution 
of the Einstein field equation, is also a solution of Eq. (\ref{2.4}) 
(for further discussion, we refer the reader to the review \cite{ReviewTEGR-Maluf}). It imediately follows that Eq. (\ref{2.4}) 
may be rewritten as \cite{ReviewTEGR-Maluf,Ref18}

\begin{equation}
\partial_\nu \left(e\Sigma^{a\lambda\nu}\right)=\frac{1}{4\kappa c}
e\, e^a\,_\mu\left( t^{\lambda \mu} + T^{\lambda \mu}\right)\,,
\label{2.5}
\end{equation}
where $t^{\lambda\mu}$, defined by

\begin{equation}
t^{\lambda \mu}=\kappa c\left(4\Sigma^{bc\lambda}T_{bc}\,^\mu-
g^{\lambda \mu}\Sigma^{bcd}T_{bcd}\right)\,,
\label{2.6}
\end{equation}
is identified as the energy-momentum tensor of the gravitational field (see \cite{ReviewTEGR-Maluf,Ref18} and references therein). 
%
It follows from Eq. (\ref{2.5}) that a \textit{continuity} (or balance) equation results \cite{ReviewTEGR-Maluf,Ref18},

\begin{equation}
\frac{d}{dt} \int_V d^3x\,e\,e^a\,_\mu (t^{0\mu} +T^{0\mu})
=-\oint_S dS_j\,
\left[e\,e^a\,_\mu (t^{j\mu} +T^{j\mu})\right]\,,
\label{2.8}
\end{equation}
in which the integration is carried out in a three-dimensional volume $V,$ bounded by a surface $S$. It should be remarked that 
on Eq. (\ref{2.8}) the tensors $t^{\lambda \mu}$ and $T^{\lambda \mu}$ appear on equal footings. On the left side of the above equation one has the time derivative of the total (gravitational and matter fields)  
energy-momentum, $P^{a}$, enclosed by a volume $V$ of the three-dimensional space \cite{Ref18,ReviewTEGR-Maluf}, 

\begin{equation}
P^a=\int_V d^3x\,e\,e^a\,_\mu (t^{0\mu}+T^{0\mu})\,.
\label{2.9}
\end{equation}
%

Using Eq. (\ref{2.5}), Eq. (\ref{2.9}) may be written in terms of $\Pi^{ai}=-4\kappa ce\Sigma^{a0i}$, 
which is the density of momentum canonically conjugate to $e_{ai}$, as  

\begin{equation}
P^a=-\int_V d^3x\,\partial_i\,\Pi^{ai}\,= -\oint_{S}dS_{i}\,\Pi^{ai}\quad.
\label{2.12}
\end{equation}
In vacuum, Eq. (\ref{2.12}) represents the gravitational 
energy-momentum vector $P^{a} = (E/c, \textbf{P})$, whereas in non-empty space-times it represents the total energy-momentum of 
the gravitational and matter fields (see Ref. \cite{ReviewTEGR-Maluf} and references therein).
Eq. (\ref{2.8}) imply, in particular, for $a=(i)=(1),(2),(3)$, that
%

\begin{equation}
{{dP^{(i)}}\over {dt}}=
-\oint_S dS_j\,\phi^{(i)j}\,,
\label{2.13.1}
\end{equation}
where
\begin{equation}
\phi^{(i)j}=4\kappa c\,\partial_\nu(e\Sigma^{(i)j\nu}) \,.
\label{2.14.1}
\end{equation}
On the left side of Eq. (\ref{2.13.1}) we have the total momentum of the gravitational and matter fields divided by time, 
what has the dimension of force. As the quantity $\phi^{(i)j}$ is a tensor 
density, its dimension will depend on the coordinates one uses. When one uses dimensionfull coordinates, $dS_j\,$ has dimension of area, what implies that the quantity $-\phi^{(i)j}$ has dimension of pressure (in this case, one has the pressure along the $(i)$-direction, over a unit area element whose normal is oriented along the $j$-direction \cite{Ref18}). From it, one can readly obtain the total pressure over the surface $S$. 
For vacuum space-times, the pressure will be just the pressure of the gravitational field. In Ref. \cite{Ref19}, the 
gravitational pressure over the event horizon of a Schwarzschild and a Kerr black holes has 
been computed. In both cases, the pressure resulted to be radially directed towards the center of the black hole.
If one uses spherical-type coordinates, for instance, then by fixing $i = r, \theta,\varphi$, $j = 1$ is associated with the radial direction. 
Hence, for this type of coordinates, one can write $\phi^{(r)1} = \phi^{(i)1}n_{i}$, where $n_{i} = (\sin\theta\cos\varphi,\,\sin\theta\sin\varphi,\,\cos\theta)$ is the unit radial vector. Thus, the tensor density along the radial direction is 

\begin{equation}
-\phi^{(r)1} = -(\sin\theta\cos\varphi \phi^{(1)1}+\sin\theta\sin\varphi\phi^{(2)1}+\cos\theta\phi^{(3)1})\,.
\label{2.15}
\end{equation}
When we consider, as we will do later, the FRW metric in non-dimensionfull $r, \theta,\varphi$ coordinates, then the quantity $\phi^{(i)j}$ 
(and so does $\phi^{(r)1}$) will just have dimension of force and so the pressure can be readily obtained from it. 
%
The expression of $-\phi^{(r)1}$ for a FRW universe will be given in the next Section and from it we will obtain the gravitational pressure over a spherical, space-like two-surface. The expression given by Eq. (\ref{2.13.1}) has been applied to the study of the thermodynamics 
of Kerr \cite{Ref19} and Reissner-Nordstr\"om \cite{karl-jf} black holes. In Ref. \cite{Ref19} the effect of the gravitational pressure 
on the Penrose process was to decrease its efficiency. In Ref. \cite{karl-jf} the investigation of the 
gravitational entropy of the black hole has led to the result that the gravitational pressure over the event 
horizon of the black hole modify the standard, Bekenstein-Hawking entropy-area relation, even if one assumes 
that the (gravitational, \textit{i.e.}, classical) thermodynamic temperature of the black hole is given by the 
(quantum) Hawking temperature. Furthermore, in the context of gravitational waves \cite{Maluf-Ulhoa2008}, it has been shown 
that gravitational pressure can be consistenly ascribed to gravitational waves.

\bigskip

\section{Gravitational Pressure in a FRW Universe}

\noindent

A homogeneous and isotropic universe is described by the (standard) 
Friedman-Robertson-Walker (FRW) metric in co-moving coordinates (see, \textit{e.g.}, \cite{MTW})

\begin{equation}
ds^{2} = -c^{2}dt^{2}+a^{2}(t)\left\{{dr^{2}\over 1 - kr^2} + r^{2}\left (d\theta^{2}+ \sin ^{2}\theta d\varphi^{2}\right)\right\}\,,
\label{2.18} 
\end{equation}
in which $k =0, 1$, or $-1$ is the curvature constant of a three-dimensional spatial hypersurface, 
$a(t)$ is the ``scale factor'' and $t$ is the ``cosmic'' time. Observations of the actual 
universe can provide information on the values of the parameters $k$ and $a$. 
In this work, we consider the the Einstein field 
equation with the cosmological term, $\Lambda\, g_{\mu\nu}$, namely, 
\begin{equation}
R_{\mu\nu}-\frac{1}{2}g_{\mu\nu}R + \Lambda g_{\mu\nu} = \frac{8\pi G}{c^4}\,T_{\mu\nu}\,, 
\label{Eq-Einstein} 
\end{equation}
not because it might be responsible for the observed cosmic speed up, but rather because the value of 
$\Lambda\,$, \textit{a priori}, cannot be set as equal to zero. It is a constant that naturally arises in the left side of the 
Einstein field equation and whose value must be determined from observations 
\footnote{In the literature, it seems to be widely remarked that $\Lambda$ should be positive due to the value of the cosmological 
observations related to the cosmic speed up. However, this conclusion can be reached not necessarily as looking for an ``explanation'' for the cosmic acceleration. It fact, when the standard cosmological model is considered, the very dynamics of general relativity leads to the conclusion that $\Lambda > 0$, since consistency cannot be achieved without the latter condition (see, \textit{e.g.}, Ref. \cite{Rindler}).}. Thus, in this work, as contrary to what is usually refered in the literature, we consider that $\Lambda$ is \textit{not} the ``source'' of dark energy, that is, it is not simply 
a term that must be considered as part of the energy-momentum of the source-term of the field equation (\ref{Eq-Einstein}), as $T_{\mu\nu}^{\Lambda}=-(\Lambda c^4/{8\pi G})\,g_{\mu\nu}$, in order to provide an explanation for dark energy. In this work, instead of 
considering that dark energy is related to vacuum energy due to a non-null cosmological constant, or some sort of it, we will see that 
one is naturally led to consider the possibily that the 
accelerated expansion of the universe might be due to the \textit{pressure of the gravitational field} itself, to which not only $\Lambda$ 
contributes, but also the matter density of the Universe. For addressing this issue we consider the definition of the gravitational energy-momentum tensor that arises in the framework of the TEGR, 
as we have reviewed on Sec. 2. 

\bigskip

Even though the gravitational field itself is not 
explicitly treated as a source in the gravitational 
field equation, it is expected that gravity itself 
must gravitate. That the gravitational field must 
contribute as source to itself (i.e., that ``gravity 
gravitates'') is due to the non-linearity of the 
gravitational field equation. That gravitational 
energy and gravitational momentum must be taken into 
account can also be seen by the gravitational field 
equation and by the Bianchi identity. In fact, the 
vanashing of the covariant derivative of the matter 
and electromagnetic energy-momentum does not expresses 
their exact conservation, but rather the gravitational 
energy-momentum must be included as well (see, \textit{e.g.}, \cite{Rindler}). 
Of course, the energy-momentum 
of the gravitational field should be properly addressed 
by means of a (gravitational) energy-momentum tensor, 
but such a quantity is generally not available in the 
usual description of general relativity. We will return 
to this issue later in the paper.

\bigskip

From Eq.s (\ref{2.18})-(\ref{Eq-Einstein}) and from the energy-momentum tensor of the cosmic fluid in its rest-frame, namely,

\begin{equation}
T_{\mu\nu}= \left(\rho c^2 + p\right)u_{\mu}u_{\nu}+ pg_{\mu\nu}\,, 
\label{fluido} 
\end{equation}
where $\rho$ and $p$ are its total masss density and pressure, respectively, and 
$u_{\mu}=(-c,0,0,0)$ is its four-velocity, it follows that  
%
%

\begin{equation}
\left(\frac{\dot{a}}{a}\right)^{2}= -\frac{kc^2}{a^2}+\frac{8\pi G\rho}{3} + \frac{1}{3}\Lambda c^{2}\,,
\label{Eq3-1}
\end{equation}
which is the Friedman equation, and

%
\begin{equation}
\frac{\ddot{a}}{a}= -\frac{4\pi G}{3}\left(\rho + \frac{3p}{c^2}\right)+\frac{1}{3}\Lambda c^{2}\,.
\label{4}
\end{equation}
Eq.s (\ref{Eq3-1}) and (\ref{4}) will be used later in this Section.

\bigskip

In order to investigate the possible effect of the gravitational pressure on a Friedman universe, we will firstly 
compute the total space-time (\textit{i.e.}, of the gravitational and matter fields) pressure over a space-like 
spherical two-surface. For this, as a first step, 
let us consider a tetrad field related to the FRW metric (\ref{2.18}). Each set of tetrad fields defines a class of reference frames in space-time. 
A set of tetrad fields adapted to an observer and which corresponds to the metric (\ref{2.18}) can be directly derived through the relationship $g_{\mu\nu}=\eta^{ab}e_{b\mu}e_{a\nu}$. Firstly, 
we recall that a tetrad field, which is the basic field variable of the TEGR, is naturally interpreted as a {\it reference frame} 
adapted to an observer in space-time (\cite{Ref25},\cite{Ref16},\cite{ReviewTEGR-Maluf}). To every observer in space-time its four-velocity along 
its world-line is given by $u^{\mu}(s)=\,e_{(0)}\,^{\mu}$. A set of tetrad fields adapted to static observers is 
achieved by imposing on $e_{a\mu}$ the following conditions: (i) $e_{(0)}\,^{i}=0\,$, which implies that $e_{(k)0}=0\,$, and (ii) $e_{(0){i}}=0\,$, 
which implies that $e_{(k)}\,^{0}=0\,$. The meaning of condition (i) is that the translational velocity of the observer is null, while condition (ii)  means that the observer (more precisely the three spatial axes of the observer's local spatial frame) 
is (are) not rotating with respect to a nonrotating frame (for details, we refer the reader to Ref. \cite{Ref16} 
and references therein). From conditions (i) and (ii), one can easily obtain the set of tetrad fields 
which corresponds to static observers and is related to the metric (\ref{2.18}). It reads 

\begin{equation}
e_{a\mu} = \left(\begin{array}{cccc}
-c & 0 & 0 & 0\\
0  & A\sin\theta\cos\varphi & a(t)r\cos\theta\cos\varphi & - a(t)r\sin\theta\sin\varphi \\
0  & A\sin\theta\sin\varphi & a(t)r\cos\theta\sin\varphi & a(t)r\sin\theta\cos\varphi \\
0  & A\cos\theta         & - a(t)r\sin\theta       & 0
\label{2.20.1}
\end{array}
\right),
\end{equation}
where $ A = {a \over \sqrt{1 - kr^2}}$. It follows that the deteminant $e=det(e^{a}_{\,\,\mu})$ is $e = cAa^{2}(t)r^{2}\sin\theta$, which 
will be needed later.

\bigskip

As a second step in order to evaluate the pressure over a spherical space-like two-surface of the FRW universe class, one 
needs to compute the quantities $\phi^{(i)1}$ given by Eq. (\ref{2.14.1}) and which appear in Eq. (\ref{2.15}). For the set of tetrad 
fields given by Eq. (\ref{2.20.1}), one can straightforwardly compute the components of the torsion tensor, 
$T_{a\mu\nu}=\partial_{\mu}e_{a\nu} - \partial_{\nu}e_{a\mu}$. The components $\Sigma^{a\mu\nu}=\eta^{ab}e_{b\lambda}\Sigma^{\lambda\mu\nu}$ 
are also straightforwardly computed, through Eq. (\ref{2.3}). The nonvanishing components of $\Sigma^{\lambda\mu\nu}$ are 

\begin{eqnarray}
\Sigma^{001}&=&\frac{1}{A^2ar}(A-a)\,,\nonumber\\
\Sigma^{110}&=& -\frac{\dot{a}}{A^{2}a}\,,\nonumber\\
\Sigma^{220}&=& -\frac{\dot{a}}{a^{3}r^2}\,,\nonumber\\
\Sigma^{330}&=& -\frac{\dot{a}}{a^{3}r{^2}\sin^{2}\theta}\,,\nonumber\\
\Sigma^{212}&=& \frac{A-a}{2A^2a^{3}r^{3}}\,,\nonumber\\
\Sigma^{313}&=& \frac{A-a}{2A^2a^{3}r^{3}\sin^{2}\theta}\,.\nonumber\\
\label{sigma}
\end{eqnarray}

\bigskip

From Eq.s (\ref{2.20.1}) and (\ref{sigma}), it follows that 

\begin{eqnarray}
\phi^{(1)1} &=& - 4\kappa c\sin^{2}\theta\cos\varphi \left(\partial_{0}(a\dot{a}){r^{2}\over c^{2}} + 1 - \sqrt{1 - kr^2}\right),\nonumber\\
\phi^{(2)1} &=& - 4\kappa c\sin^{2}\theta\sin\varphi \left(\partial_{0}(a\dot{a}){r^{2}\over c^{2}} + 1 - \sqrt{1 - kr^2}\right),\nonumber\\
\phi^{(3)1} &=& - 4\kappa c\sin\theta\cos\theta \left(\partial_{0}(a\dot{a}){r^{2}\over c^{2}} + 1 - \sqrt{1 - kr^2}\right).
\label{2.24}
\end{eqnarray}
Inserting now the above relations into Eq. (\ref{2.15}) and performing its integration in the angular variables, one obtains that 
(recall that $x^{0}=t$)

\begin{equation}
p(r)= {c^{4}\over G}\left\{1 + \left[\left(\frac{\dot{a}}{a}\right)^{2}+\frac{\ddot{a}}{a}\right]{r^{2}a^{2}\over c^{2}}-\sqrt{1-kr^{2}}\right\}\,.
\label{rad-press}
\end{equation}
This result implis that the quantity $p(r)$, which has dimension of force, is isotropic, what complies with 
the symmetry of the FRW universes.

\bigskip

Let us now consider a spatially flat universe, what corresponds to take $k = 0$, whose direct evidence comes from the anisotropies of the 
cosmic microwave background radiation (see, \textit{e.g.}, \cite{Rindler}). In this case, Eq. (\ref{rad-press}) reduces to
\begin{equation}
p(r)= {c^{2}\over G}\left[\left(\frac{\dot{a}}{a}\right)^{2}+\frac{\ddot{a}}{a}\right]r^{2}a^{2}\,.
\label{5a}
\end{equation}
Substituing into the latter its corresponding terms which appear in Eq.s (\ref{Eq3-1}) (taking $k=0$) and (\ref{4}), one obtains  

\begin{equation}
p(r)= \left(\frac{4\pi}{3}\rho c^{2} - 4\pi p +\frac{2}{3}\frac{\Lambda c^{4}}{G}\right)r^{2}a^{2}\,.
\label{7-2}
\end{equation} 
Recall that $p$, on the right-hand side of (\ref{7-2}), is the pressure of matter and, of course, it should not be confused with $p(r)$ which figures 
on the left side of the above equation. From the quantity (\ref{7-2}) one can obtain the pressure, $P$, over a spherical surface of proper radius $ar$ 
as $P=\frac{p(r)}{4\pi a^{2}r^{2}}$. Thus, one is left with

\begin{equation}
P= \frac{1}{3}\rho c^{2} + \frac{1}{6\pi}\frac{\Lambda c^{4}}{G} -p\,.
\label{pressao}
\end{equation} 
From this result, it follows that $P$ does not depend on the proper radius, $ar$, of the surface. 
Thus, the pressure $P$ is not only isotropic, as we have earlier verified, but also it is spatially homogeneous. Hence, 
it complies with the fact that the universe is homogeneous and isotropic (recall that we have chosen a set of static, co-moving observers to deal with, according to the set of tetrad fields given by Eq. (\ref{2.20.1})). Now, in order one obtains some idea of the effect of the pressure given by Eq. (\ref{pressao}), and possibily on the expansion of the universe, let us consider the \emph{apparent horizon} of the universe (see, \textit{e.g.}, \cite{Faraoni},\cite{Hayward} and references therein). As opposed to an event horizon, an apparent horizon is not a global one, but rather it is defined (quasi-)locally (for details, we refer the reader to, \textit{e.g.}, Ref. \cite{Hayward}). Besides, the apparent horizon always exists in all FRW space-times, as opposed to event and particle horizons (see, \textit{e.g.}, \cite{Faraoni} and \cite{Rindler}). 
As we are considering a spatially flat FRW universe, the apparent horizon coincides with the ``Hubble horizon'' 
\footnote{The location of the apparent horizon can be found directly if it is characterized as 
a momentarily stationary light-front in terms of its area coordinate $\tilde{r}_{AH}=ar_{AH}$ ($r_{AH}$ is the 
radius in co-moving coordinates) \cite{Rindler}. The apparent horizon is located at  
$\tilde{r}^{2}_{AH}=\frac{a^{2}c^2}{\dot{a}^{2}+kc^{2}}$ (see, \textit{e.g.}, \cite{Faraoni} and \cite{Rindler}), that is, at the radius $
\tilde{r}_{AH} = {c \over \sqrt{H^{2} + kc^2/a^{2}}}\,$,
where $H = \dot {a}/a$ is the Hubble parameter. Thus, for a spatially flat FRW universe the apparent horizon coincides with the Hubble horizon. We also note that in the case of a spatially flat FRW universe the area distance (or area coordinate) coincides with the proper distance.}. 
We also remark that, as compared to event horizons, apparent horizons seem to be more suitable for formulating a consistent thermodynamics of cosmological space-times 
(see,\textit{ e.g.}, \cite{Faraoni} and references therein). From now on, one can consider the pressure (\ref{pressao}) as simply the pressure over the apparent horizon (or Hubble horizon) of a spatially flat FRW universe. 

\bigskip

As the present universe is in an era in which radiation pressure is negligible and matter is non-relativistic (``cold matter''), we 
take $p=0$ in Eq. (\ref{pressao}). Thus, one is left with

\begin{equation}
P = \left(\frac{1}{3}\rho + \frac{1}{6\pi}\frac{\Lambda c^{2}}{G}\right)c^{2}\,.
\label{pressao2}
\end{equation}
Therefore, in the case of a spatially flat FRW universe and for the actual era, it results that, for $\Lambda > 0\,$, 
the pressure $P$ is positive, what implies that $P$ is 
\textit{outwardly} directed over any spherical, space-like two-surface of the flat FRW universe (thus, in particular, over the Hubble horizon). 
Consequently, we could think of the pressure $P$ as being responsible for the 
expansion of the universe as a whole, as the latter evolves. Thus, as according to observations the actual universe is under accelerated expansion, the 
latter could be simply the result that the Hubble horizon is being pushed by the pressure of the very gravitational field. We remark that the classical electromagnetic field is endowed with pressure, or, more generally, with stress-energy-momentum. Hence, one would expect the same to hold 
for the classical gravitational field. In the following, we will discuss more on the possibility that the pressure of the gravitational field 
could be responsible for the actual acceleration of the universe.  

\bigskip 

%
%
%
In view of the result given by Eq. (\ref{pressao2}), for later reference let us define the quantity 

\begin{equation}
\rho_{eff}\equiv\,\frac{1}{3}\rho + \frac{1}{6\pi}\frac{\Lambda c^{2}}{G}\,,
\label{dens-ef}
\end{equation}
with which Eq. (\ref{pressao2}) can be simply rewritten as

\begin{equation}
P = \rho_{eff}c^{2}\,.
\label{pressao3}
\end{equation} 

\bigskip
%
%
%
%

Firstly, from Eq. (\ref{pressao}) or Eq. (\ref{pressao2}), we note that the gravitational pressure is not exclusively due 
to the $\Lambda$-term. Even in the absence of the $\Lambda$-term, a nonvanishing gravitational 
pressure would result. Also, from Eq. (\ref{pressao}), as a particular case, it follows that for an empty de Sitter universe 
(for which $\rho = p =0$), the total pressure reduces simply to $P=\Lambda c^{4}/6\pi G$, which is independent of the cosmic time.  
Furthermore, still according to Eq. (\ref{pressao}), in the epochs of the universe when $p$, the pressure of the cosmic fluid 
(here, we assume it takes only positive values, \textit{i.e.}, we do not consider dark energy), is non-negligible the total pressure, $P$, 
depends on the value of $p$. For instance, in the radiation-dominated era, one has $p=\rho c^{2}/3$, what implies that $P$ achieves the value which is 
precisely that for the empty de Sitter universe. And for negligible $p$, as is the 
case of the present universe, it follows that $P$ is given by Eq. (\ref{pressao2}). Hence, the total, space-time pressure 
over any spherical space-like two-surface of the universe achieves its greatest value in the present epoch. Thus, 
the total, space-time pressure, $P$, increasead as the universe expansion occured. Since the pressure of matter has descreased during 
the expansion of the universe to the present era, we conclude that the pressure of the gravitational field has increased. Hence, the 
expansion of the universe is related to the pressure of its very gravitational field.

\bigskip

We will now relate our expression (\ref{pressao3}) with observational data of the actual epoch of the universe. 
By expressing Eq. (\ref{dens-ef}) 
in terms of the dimensionless parameters $\Omega_{m0}\equiv\frac{8\pi G\rho_{m}}{3H_{0}^{2}}$ and $\Omega_{\Lambda 0}\equiv\frac{\Lambda c^{2}}{3H_{0}^{2}}$, we have 

\begin{equation}
\rho_{eff}= \frac{1}{3}\rho_{cr}\left(\Omega_{m0}+4\Omega_{\Lambda 0}\right)\,,
\label{dens-ef2}
\end{equation}
where $\rho_{cr} =\frac{3H_{0}^{2}}{8\pi G}$ is the critical density. The index zero in all the latter quantities means that 
they are evaluated in the present time. The above expression will be used in the next Section.

\bigskip

\section{Gravitational Pressure and the Thermodynamics of the Universe}

\noindent

As temperature and entropy seem to be intrinsic properties one can ascribe to a horizon, let us assume that a temperature and an entropy  
are associated to the apparent horizon of a spatially flat FRW universe (again, we note that in this case the apparent and 
Hubble horizons coincide with each other) and that they are given by \footnote{The same assumption is made in Ref. \cite{smoot}, but in another context regarding an explanation for the cosmic 
speed up.} 

\begin{equation}
T=\frac{\hbar}{\kappa_B}\frac{H}{2\pi}\,
\label{temperature}
\end{equation}  
and 

\begin{equation}
S=\frac{\kappa_B c^3}{G\hbar}\frac{A}{4}\,,
\label{entropy}
\end{equation}  
where $A$ is the area of the horizon, given by $A=4\pi\tilde{r}^{2}_{AH}\,$, with $\tilde{r}_{AH}=c/H$ being the radius of the Hubble horizon, which 
coincides with the radius of the apparent horizon, in the case of a spatially flat universe (see footnote 2). 

\bigskip

Now, let us assume that the thermodynamic relation $TdS=dE + PdV$ holds. In the latter, $E$ is the total energy of the space-time, due to the 
gravitational and matter fields. It is given by the zero-component of the total (of gravitational and matter fields) 
energy-momentum vector $P^a$ (see Eq.(\ref{2.12})), that is, $E=cP^{(0)}$. The latter quantity has been computed to the case of a 
FRW universe in Ref. \cite{Maluf-IJMPD-EnergyFRW} and it resulted that $E=(c^{4}/G)ar(1-\sqrt{1-kr^{2}})$ (the same result was 
obtained, by a totally different method in Ref. \cite{Chen-Liu-Nester2007}). Hence, in the case of a spatially flat FRW 
universe ($k=0$) one has $E=0$, what means that the energy of the gravitational and matter fields cancel each other, thus corresponding to a 
spatially flat universe. Hence, it follows that the thermodynamic relation $TdS=dE + PdV$ is reduced to 
\begin{equation}
TdS = PdV\,.
\label{thermo-relation}
\end{equation}
Since we have found, considering the actual universe with ``cold matter'' and spatially flat (see Eq. (\ref{pressao2})), that $P\,>\,0$, it follows from the above relation that $dS\,>\,0$ in an expanding spatially flat universe, that is, the gravitational entropy increases as the actual universe 
expands. 

\bigskip

Now, from Eq. (\ref{entropy}) and Eq. (\ref{thermo-relation}) we have 

\begin{equation}
T\,\frac{\kappa_B c^2}{G\hbar} = \frac{2}{H}\,P\,.
\label{thermo-relation-2}
\end{equation}
And after replacing Eq.s (\ref{pressao3}) and (\ref{dens-ef2}) into the above equation we are left with 

\begin{equation}
T\,\frac{\kappa_B}{\hbar} = \frac{H}{2\pi}\,\left(\frac{1}{2}\Omega_{m0}+2\Omega_{\Lambda 0}\right)\,.
\label{thermo-relation-3}
\end{equation} 
Finally, taking into account Eq. (\ref{temperature}), we are left with
\begin{equation}
\Omega_{m0}+4\Omega_{\Lambda 0} = 2\,.
\label{relation-cosmo-param}
\end{equation} 

Since $\Omega_{m0}\approx 0.3$, we arrive at 
\begin{equation}
\Omega_{\Lambda 0}\approx 0.425\,.
\label{estima-param}
\end{equation} 
Thus, it results that 
\begin{equation}
\Omega_{m0}+\Omega_{\Lambda 0}\approx 0.725\,<\,1\,,
\label{relation-cosmo-param-2}
\end{equation} 
what is expected, since no contribution due to the stress-energy-momentum of 
the gravitational field was explicitely included as a source-term in Einstein field equation and, thus, in Friedman equation (Eq.(\ref{Eq3-1})). 
Hence, no contribution of the very gravitational field has been included in the cosmic budget. However, as the interaction of the gravitational field with itself contributes to its own source, such a contribution is physically expected to play a role. 
From Eq. (\ref{relation-cosmo-param}) and Eq. (\ref{dens-ef2}) it follows that 
\begin{equation}
\rho_{eff}= \frac{2}{3}\rho_{cr}\,,
\label{dens-ef-dens-cric}
\end{equation}
what implies, from Eq.(\ref{pressao3}), that 

\begin{equation}
P = \frac{2}{3}\rho_{cr}c^{2}\,,
\label{dens-ef-dens-cric-final}
\end{equation}
a value that is very close to the one usually ascribed to dark energy. 

\bigskip

We remark that in Ref. \cite{smoot} the acceleration of the 
universe is due to an entropic force whose corresponding pressure is given by $-\frac{2}{3}\rho_{cr}c^{2}$. The minus sign just means that the 
entropic force points in the direction of increasing entropy (\textit{i.e.}, to the Hubble horizon). We point out that 
this result was derived, in Ref. \cite{smoot}, based on the Friedmann equation and on the assumption of that one can ascribed the temperature (\ref{temperature}) and the entropy (\ref{entropy}) to the Hubble horizon due to the information holografically stored there. In Ref. \cite{smoot}, no use of the basic relation (\ref{thermo-relation}), \textit{i.e.}, of the firt law of thermodynamics, was done in deriving the pressure from the entropic force. Our idea to explore the thermodynamics of the universe based on (\ref{thermo-relation}), in the context of the TEGR, was prior to our knowledge of the Ref. \cite{smoot}. Nevertheless, it is remarkable that we arrived essentially at the same expression of Ref. \cite{smoot}, by means of a different approach. 

\bigskip

Now if we consider that the effect of the gravitational energy-momentum must be taken into account in the cosmic budget, we must have 
\begin{equation}
\Omega_{m0}+\Omega^{\prime}_{0} = 1\,,
\label{sum-param2}
\end{equation}
instead of (\ref{relation-cosmo-param-2}), where $\Omega^{\prime}_{0} = \Omega_{\Lambda 0} + \Omega_{g0}\,$, with $\Omega_{g0}$ being 
the contribution due to the \textit{ pressure of the gravitational field}. Taking into account Eq. (\ref{sum-param2}) and 
the value for $\Omega_{m0}$ and that given by Eq. (\ref{estima-param}), we are left with $\Omega_{g0}\approx 0.275$.   

\bigskip

We remark that  
$\Omega_{g}=1-(\Omega_{m}+\Omega_{\Lambda})=\frac{\rho_{cr}-(\rho_{m}+\rho_{\Lambda})}{\rho_{cr}}$ is a function of time, 
by means of the cosmic time dependence of $\rho_{m}$ and of $\rho_{cr}$. We interpret the time dependence of 
$\Omega_{g}$ as a result that the gravitational stress-energy-momentum changes with time as the universe expands, since we have found 
that the gravitational pressure is \textit{outwardly} directed over the Hubble horizon. Hence, the gravitational pressure might be responsable 
for the observed accelerated expansion, instead of a mysterious, dark energy. As a result, the measurements that show 
evidence of the cosmic speed up would be related not only to the cosmological constant, but also to the pressure of 
gravity itself, which, due to its dependence with the cosmic time, is a dynamical quantity.
  
\bigskip

\section{Final Remarks}

\noindent

In this work, we have examined the possibility that the issue of the accelerated expansion rate of the actual universe might be related to a basic concept of one can ascribed to the very gravitational field, namely the \textit{gravitational pressure}. The latter naturally 
arises in the framework of the \textit{Teleparallelism Equivalent to General Relativity} (TEGR) \cite{Ref18},\cite{ReviewTEGR-Maluf}.
Instead of the usual approach for the description of the acceelerated expansion of the universe, in the present work we have not 
assumed that the cosmological constant would be responsible for the dark energy, but rather we have applied the definition of the stress-energy-momentum of the gravitational and matter fields, which is directly identified from the field equation of the TEGR, which in turn is 
equivalent to Einstein field equation. It is interesting to explore the stress-energy-momentum of the gravitational field in order to describe fundamental aspects of gravity. 
From the definition given by Eq. (\ref{2.13.1}), it is possible to define an expression for the total pressure of the space-time,\textit{ i.e.}, of the 
gravitational and matter fields. We remark that the result (\ref{2.13.1}) steems from a balance (or continuity) equation [see Eq. (\ref{2.8})], 
which in turn comes from the field equation of the TEGR [see Eq. (\ref{2.5}) or Eq. (\ref{2.4})]. We also stress that the gravitational and matter 
stress-energy-momentum tensors take place in the field equation of the TEGR on the same footing. Such a property is not present in the description 
of gravitation as formulated in General Relativity.

\bigskip

We have computed the total space-time pressure (\textit{i.e.}, of gravitational and matter fields) over a spherical, space-like 
two-surface of the FRW universe, for any value of the curvature index, $k$. In particular, in 
the case of a spatially flat FRW universe, it resulted that the pressure is independent of the radius of the surface 
[see Eq. (\ref{pressao})], what complies with the homogeneity and isotropy of the universe, as well as with the choice we have made use of the tetrad fields adapted to co-moving observers [see Eq. (\ref{2.20.1})]. 
In order we could address the possible effect of the pressure of the gravitational field on the present 
expansion of the universe as a whole, we have focused attention on the surface of the cosmological apparent horizon, which 
for the spatially flat case, is equal to the Hubble horizon. We have considered the case of a spatially flat FRW universe, as the data on the cosmological observables is usually considered to favors it. Hence, as in the present universe the matter is cold, the pressure of the 
space-time is entirely due to the gravitational field. We have found that the gravitational pressure is positive (see Eq. (\ref{pressao2})) 
and thus it is \textit{outwardly} directed over the cosmological apparent (or Hubble) horizon. Consequently, one can think of the 
gravitational pressure as acting to cause the repulsive effect which would be responsible for the present accelerated expansion of the universe. 
Hence, the universe could be under accelerated expansion basically due to the effect of the pressure of its own gravitational field, rather than due to some exotic, totally unkown sort of energy (\textit{i.e.}, the dark energy). 
We recall that $P,$ given by 
Eq. (\ref{pressao3}), stems from the total stress of the gravitational and matter fields, as defined in the realm of the TEGR. 
In summary, our result (\ref{pressao3}) takes into account not only the usual contributions of the matter fields to the dynamics 
of the universe but also the contribution of the gravitational field itself (even in the absense of the $\Lambda$-term). In other words, the contribution of the pressure of the gravitational field might really be playing be responsible for the expansion of the universe. 
As far as we know, that the pressure of the gravitational field, as derived from the stress-energy-momentum of the gravitational field itself, could play a role in the actual accelerated expansion of the universe has not been addressed in the literature.

\bigskip

We recall that it is widely argued in the literature that as the present universe is accelerated ($\ddot{a}>0$), it follows 
from Eq. (\ref{4}) that $\rho + 3p/c^{2}\,<\,0$. This implies that $p$ would be negative, and then  
an exotic sort of energy would have to be taken into account (the so-called dark energy) accordingly. Nevertheless, one should bear in mind 
that this argument is based on the \textit{Newtonian interpretation of the Eq. (\ref{4})} (see, \textit{e.g.}, 
Sec. 18.2.E of Ref. \cite{Rindler}). 

\bigskip

During the development of this work, we have found the interesting proposal of an alternative interpretation for dark energy discussed in Ref. \cite{smoot}. In the latter, the authors have taken into account the entropy and temperature intrinsic to the horizon of a spatially flat 
FRW universe. Those entropy and temperature were considered to be due to the information that is holographically stored on the apparent   
horizon (in this case, given by the Hubble radius). As argued in Ref. \cite{smoot}, the cosmic acceleration would be due to an \textit{entropic} 
force that would arise from the information stored on the horizon surface. We have noted that in the context of our approach 
one can naturally ascrib entropy to the gravitational field by means of the fisrt law of thermodynamics (which is fundamentally 
a statement of conservation of mass-energy), namely, $TdS = dE + PdV$, where $T$ we have identified with the temperature of the apparent horizon 
(in general, $T$ is considered to be defined 
by the surface gravity of a horizon), $S$ is the entropy of the gravitational field related to that surface (there is no contribution to 
the total entropy from the adiabatic cosmic fluid), $E$ is the total energy of gravitational and matter fields (which is zero 
for a spatially flat universe), and $P$ is the total, space-time pressure due to the matter and gravitational 
fields (for the actual universe, $P$ is purely from gravitational origin, since matter is considered to be ``cold''. 
In this way, the thermodynamics of the apparent horizon of the FRW universe is thus simply established in this way. 

\bigskip

Finally, we would like to stress that from (i) the definition, in the framework of the TEGR, 
of the total stress-energy-momentum of the gravitational and matter fields (what ensured the total energy and the total pressure 
of the space-time), from (ii) the first law of thermodynamics, and from (iii) the atribution of temperature and entropy to the 
apparent horizon of a spatially flat FRW universe we have obtnained that the pressure (in this case, a \textit{purely grativitational} pressure, 
since for the actual universe the matter is pressureless) is given 
by $P=(2/3)\,\rho _{cr} c^2\,$, what is the same, except for the opposite sign, as that obtained in Ref. \cite{smoot}. In the latter, 
the negative sign corresponds to the fact that the pressure is a tension over the horizon surface and the corresponding 
force is an entropic force. It is remarkable that two different aproaches, which are both model-independent, have led to equivalent 
results relating the issue of the cosmic speed up to a simple general property of the very gravitational field, namely, its pressure. 

\bigskip

We have found that the when the matter is cold (as is in the present universe) it results that the total pressure (totally due to gravity) is outwardly directed over any spherical, spatial two-surface. This has led us to propose that the cosmic speed up of the actual universe might be an effect of the pressure of the very gravitational field. In this case, there would be no need to resort to the idea of a totally unknown and arbitrary \textit{dark energy}. Interestingly, we note that modified teleparallel gravity also allows an alternative understanding of the cosmic acceleration problem, without resorting to the idea of dark energy (see, for instance, Ref.s \cite{Modif-F(T)} and references therein). However, our approach is totally different and we do not even consider $f(T)$ theories,\textit{i.e.}, modified teleparallel gravity. Furthermore, in Ref.s \cite{Modif-F(T)} the idea of gravitational pressure is not considered at all.

\end{document}